\documentclass[aps,
floatfix,notitlepage,
longbibliography
]{revtex4-1}

\usepackage{hyperref}
\usepackage{amsmath}
\usepackage{graphicx}

\newcommand\Rey{\mbox{Re}}
\newcommand\beq{\begin{equation}} 
\newcommand\eeq{\end{equation}} 
\newcommand\bea{\begin{eqnarray}} 
\newcommand\eea{\end{eqnarray}} 

\begin{document}

\title{Transition to turbulence in shear flows}

\author{Bruno Eckhardt}
\affiliation{Fachbereich Physik, Philipps-Universit\"at Marburg, 35032 Marburg, Germany}

\begin{abstract}
Pipe flow and many other shear flows show
a transition to turbulence at flow rates for which the laminar profile is stable 
against infinitesimal perturbations.
In this brief review the recent progress in the understanding of this transition
will be summarized, with a focus on the linear and nonlinear states that drive
the transitions, the extended
and localized patterns that appear, and on the spatio-temporal dynamics
and their relation to directed percolation.
\end{abstract}

\nopagebreak

\maketitle
\nopagebreak

\section{Introduction}
The ways in which flows become turbulent can broadly be divided into 
two groups. One group contains all flows where the laminar profile shows 
a linear instability as the flow rate increases. A subsequent cascade of
instabilities results in ever more complicated dynamics, which in 
spirit, though not in detail, reflects Landau's descriptions of the transition to 
turbulence \cite{Landau:1944tw}. Fluids sheared 
between independently rotating 
concentric cylinders (Taylor-Couette flow) or heated from below (Rayleigh-Benard flow)
follow this route and the 
linear and  nonlinear properties of the patterns that form have been explored in considerable 
detail \cite{Busse:1978bd,Koschmieder93,Cross:1993ab,Aranson:2002vf,Manneville14}.

\begin{figure}[ht]
\begin{center}
\includegraphics[width=0.8\textwidth]{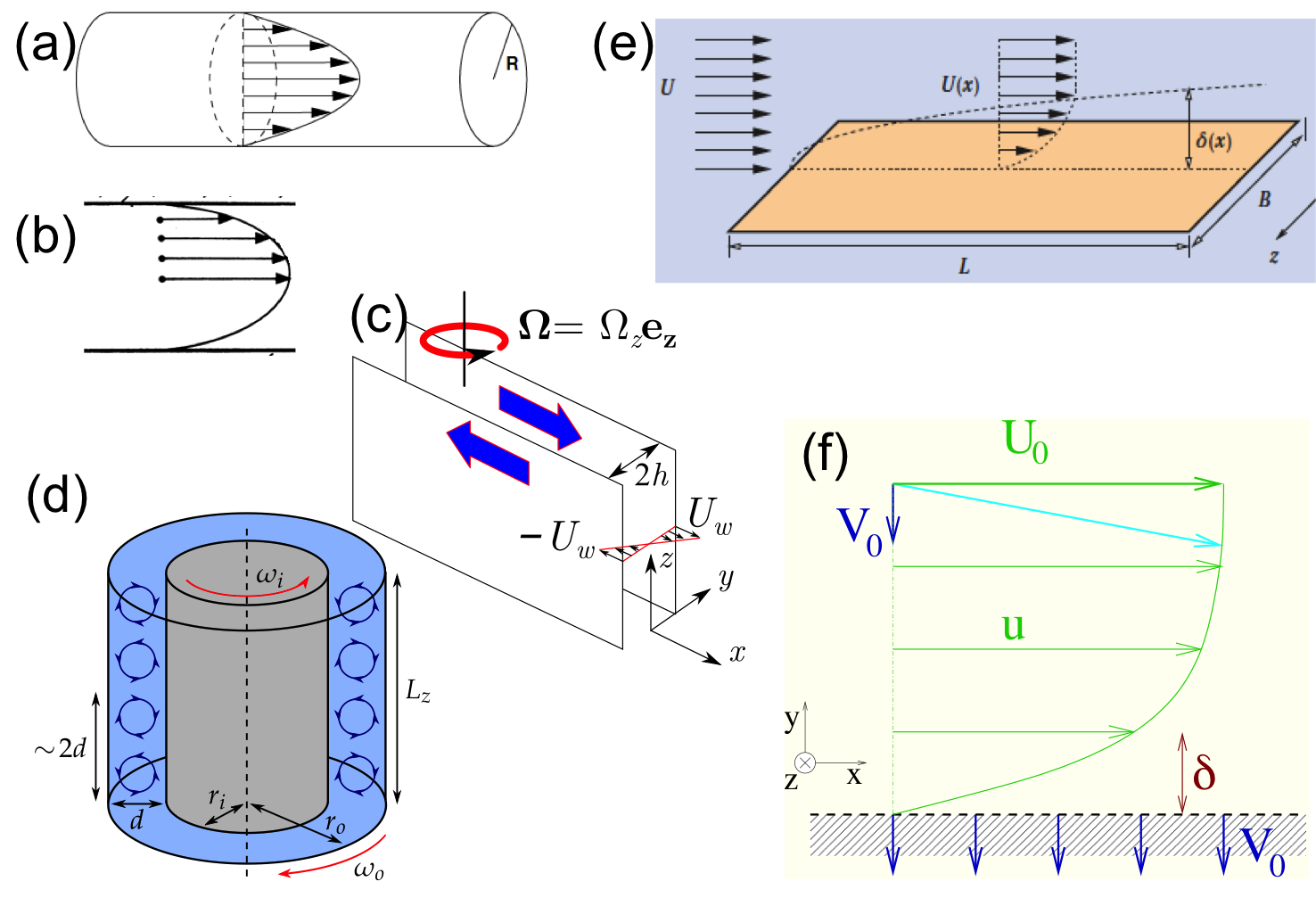}
\end{center}
  \caption{(Colour online) Examples of flows where the transition to turbulence is induced
  by sufficiently strong perturbations and is not connected with a linear instability of the laminar profile. The left column
  shows  internal flows bounded by walls on two sides, from top to bottom:  (a) pressure driven flow in a cylindrical pipe of 
  circular cross section (Hagen-Poiseuille flow); 
 (b)  the related pressure driven flow between parallel plates (plane Poiseuille flow);
 (c) shear flow between parallel walls moving relative to each other on a rotating table (rotating plane Couette flow); 
 (d)  flow between two independently
  rotating  concentric cylinders (Taylor-Couette flow). The right column shows external flows bounded by a wall on
  one side only: (e) flow over a flat plate where the boundary layer increases with distance from the edge (Blasius) and 
  (f) the suction  boundary layer where a cross-flow maintains a parallel boundary layer of constant thickness. Of the flows
  shown here, plane Poiseuille flow, Taylor-Couette, and both boundary layer flows  also have linear instabilites
  that become relevant in certain parameter ranges. The focus here is on transitions in parameter ranges 
  without linear instabilites. Coordinates are usually chosen such that $x$ points in the flow direction, $y$ in the
  direction with the shear and $z$ in the spanwise neutral direction.
  }
  \label{fig:shear_flows}
\end{figure}

The second group has pressure driven flow in a pipe as a paradigmatic example 
\cite{Reynolds:1883ty,Reynolds:1883wv}, but also includes plane Couette flow or
boundary layers and several other cases shown in Fig. \ref{fig:shear_flows}.
They all share the feature that turbulence appears for parameters where the laminar
profile is still stable. Their transition typically requires finite amplitude perturbations,
it shows a sensitive dependence on initial conditions, and there are no simple 
patterns above the onset \cite{Grossmann:2000}. 
Moreover, the flows show a remarkable variety of spatio-temporal dynamics
near onset \cite{Prigent:2002fq,Moxey:2010es,Duguet:2010dv,Sano:2016kh,Lemoult:2016ks}.

In the last two decades, various elements for an explanation of the transition in these flows have identified
and explored, so that we now have a framework in which pipe flow and other flows can be approached.
The following sections provide a brief survey of key concepts such as exact coherent states (ECS),
secondary bifurcations and the formation of turbulent transients and spatio-temporal patterns, 
and the connection to phase transitions of the directed percolation type.
Various other aspects of the transition are discussed in 
\cite{Kerswell:2005ir,Eckhardt:2007ix,Eckhardt:2007ka,Willis:2008bb,Mullin:2011jm,Kawahara:2012iu,Barkley:2016wn,Manneville:2017bc,Gibson:2009kp,Song:2014xy}.

In the following sections, we describe the non-normal amplification in the
linearized dynamics (section 2), finite amplitude subcritical bifurcations that
give rise to exact coherent structures around which turbulence can form (section 3), 
the presence of localized exact coherent structures (section 4), the long-time dynamics
of localized structures and their relation to directed percolation (section 5). 
A few concluding remarks and an outlook on open issues are given in section 6.

\section{Linear approaches}

\begin{figure}[ht]
\begin{center}
\includegraphics[width=0.9\textwidth]{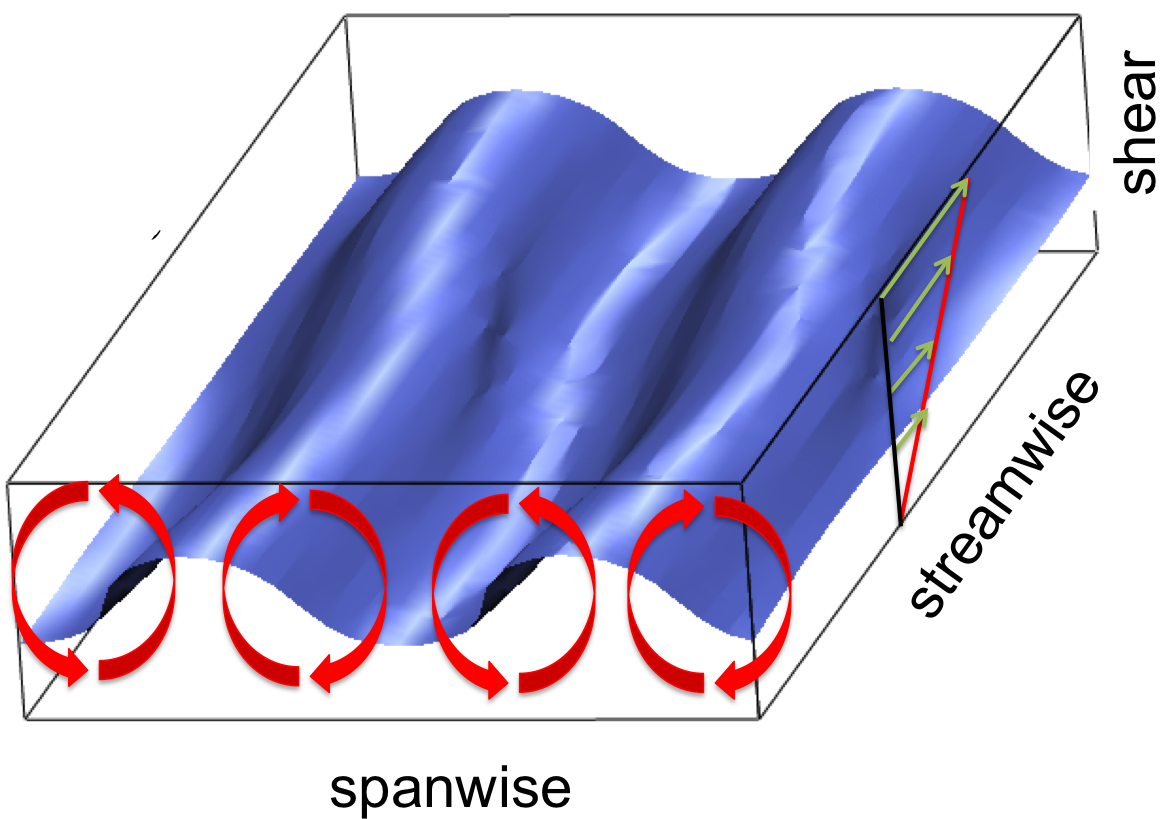}
\end{center}
  \caption{(Colour online) 
A sketch of the building blocks of turbulence in shear flows: vortices that are predominantly oriented
in the streamwise direction mix the fluid across the shear and drive a streak, a modulation in the
streamwise velocity. The blue surface shows the positions where the streamwise velocity equals
the mid-plane value of the laminar shear flow. The modulations in height are due to the vortices
and the spanwise modulations indicate that the structures are fully three-dimensional.}
   \label{fig:ECS_pcf}
\end{figure}

The linearization of the Navier-Stokes equation around a background shear
$\mathbf{U}_0$ reads
\beq
\partial_t \mathbf{u} + (\mathbf{U}_0 \cdot \nabla) \mathbf{u} + 
(\mathbf{u} \cdot \nabla) \mathbf{U}_0+ \nabla p = \nu \Delta \mathbf{u} \,.
\eeq
Asymptotic stability requires that all perturbations decay eventually. This does not
imply that perturbations decay monotonically: 
they can transiently be amplified beyond their initial amplitude before they
disappear eventually.
Typically, the structures that are most efficient in extracting energy from the
background shear are downstream vortices which then drive the formation
of modulations in the downstream velocity, so-called streaks. This interaction
provides the building blocks for the nonlinear states  as indicated
in Fig. \ref{fig:ECS_pcf}.

The significance of vortices and streaks can be studied in a simple model
\cite{Eckhardt:2003cy}. 
Consider a shear flow with $x$ pointing in the direction of flow, $y$ in the normal direction
with the shear and $z$ in the spanwise direction. Then the background velocity field 
with a linear shear is given by 
$\mathbf{U}_0=S y \mathbf{e}_x$. Adding as a perturbation 
 the superposition of a vortex $\omega$ and a streak $\sigma$,
 \beq
\mathbf{u} = \omega(t) 
\begin{pmatrix}
0\cr 
\beta \sin{\alpha y} \sin{\beta z}
\cr
 \alpha \cos{\alpha y} \cos{\beta z}
\end{pmatrix} 
+ \sigma(t) 
\begin{pmatrix}
- \beta \sin{\alpha y} \sin{\beta z}
\cr 0 \cr 0
\end{pmatrix}\,.
\eeq
The resulting equations for the perturbation become 
\beq
\partial_t 
\begin{pmatrix}
\sigma \cr \omega
\end{pmatrix}
=
\begin{pmatrix}
- \Lambda & S \cr
0 & -\Lambda
\end{pmatrix}
\begin{pmatrix}
\sigma \cr \omega
\end{pmatrix}
\eeq
with the damping constant $\Lambda=\nu \left(\alpha^2 + \beta^2\right)$.
For the initial condition $(\sigma, \omega)(0)=(0,1)$ of a vortex but no 
streak, the time evolution becomes 
\beq
\begin{pmatrix}
\sigma(t) \cr
\omega(t)
\end{pmatrix}
= \begin{pmatrix}
S t e^{-\Lambda t}\cr
e^{-\Lambda t}
\end{pmatrix} \,.
\eeq
The linear increase in the streak is a consequence of the ability of the vortex to extract energy
from the background shear flow. However, the flow will ultimately decay: all perturbations that
are translationally invariant in the downstream direction will decay \cite{Moffatt:1990fb}. 
The flow is therefore asympotically stable (in linear approximation).

More stringent than asymptotic stability is the requirement of energy stability, i.e. that the energy 
content in all perturbations  has to decay monotonically,  $d/dt \int \left( \mathbf{u}^2/2 \right) dV < 0$,
Ref. \cite{Joseph:1969th}.
For the above model, energy stability is possible for $S<2\Lambda$. If the shear $S$ becomes sufficiently 
strong,  the vortices are able to deposit enough energy in the streaks to given a transient increase in the the total energy.
Detailed linear stability analyses of various shear flows and their transient
amplification that take into account the correct boundary conditions 
are given in \cite{Schmid99,Boberg:1988uj,trefethen:1993ab,Reddy:1993up}.

A very interesting problem arises when the flows are studied in the presence of noise,
\cite{OrtizdeZarate:2008dm,OrtizdeZarate:2009kh,Sengers:2010jz,OrtizdeZarate:2011kk,OrtizdeZarate:2012dq}.
As shown for a simple model in Ref. \cite{Pausch:2015gr}, the noisy forcing of vortices
can deposit more energy in the streaks than the noisy forcing, thus explaining the
prominence of streaks in noisy shear flows. A dramatic consequence of this amplification 
has been studied by Luchini
\cite{Luchini:2010wf,Luchini:2017kl}:
The Navier-Stokes equations are continuum equations that average over
microscopic molecular fluctuations. The microscopic constituents give rise to thermodynamic
fluctuations around the continuum fields that are always present and increase with temperature
\cite{landau1969statistical}. 
If there are processes that amplify the fluctuations by $O(\Rey)$, it is possible that 
thermodynamic fluctuations are amplified so much that they cross the threshold for the
transition to turbulence. 
Indeed, Luchini's estimates in Ref. \cite{Luchini:2010wf,Luchini:2017kl} show
that thermal fluctuations are big enough to cause transition in 
boundary layers for $\Rey\approx 6\cdot10^6$. Similarly, 
Meseguer and Trefethen \cite{Meseguer:2003kq} 
noted that non-normality also causes a catastrophic loss of 
numerical accuracy that requires special precautions for
linear stability analyses of pipe flow beyond $\Rey\approx 10^7$.
These values are only a little bit higher than the highest Reynolds number of order $O(10^5)$
up to which laminar pipe flow has been realized in experiments
\cite{pfenniger1961,Draad:1998ua}.

\section{Nonlinear effects and bifurcations}
The transient amplification of the streaks
by the vortices by itself will not be sufficient to drive turbulence,
as all structures that are translationally invariant in the
downstream direction are bound to decay \cite{Moffatt:1990fb}. However, the
built up of streaks holds the seed for the formation of a linear instability near the
inflection point in the downstream profile in the shearing region 
between the streaks \cite{Waleffe:1997va}. That instability will introduce 
spanwise modulations in the downstream direction, i.e. rather than being straight 
in the downstream direction, the vortices begin to show some modulation,
and the modulated waves are capable of sustaining themselves. 
An example of the modulated exact coherent structures in plane Couette 
flow is shown in Fig. \ref{fig:ECS_pcf}.

On a technical level, the appearance of 3-d structures can be 
shown by embedding the flow into a family of flows
where one parameter is the background flow (as measured by its 
Reynolds number) and the other parameter is a forcing that sustains
the streaks (or the vortices, either way is fine). Once the instability 
has set in and the structures that break the downstream translational
invariance have been identified, the forcing of the streaks is reduced
and 3d exact coherent structures for the initial field are obtained.
This technique was initially used by Clever and Busse \cite{Clever:1997tq} for 
Rayleigh Benard convection, where they applied a temperature difference to 
drive convection rolls, and by Nagata in rotating plane Couette flow \cite{Nagata:1990ae}, 
where the second parameter in addition to shear was rotation. 
For pipe flow, Ref. \cite{Faisst:2003hd}
used an artificial body force mimicking vortices, and Ref. \cite{Wedin:2004ey}
used a body force derived from linear modes that were maximally amplified.
In general, the path to exact coherent structures does not seem to depend very much on the 
details of the forcing as long as it drives vortices and streaks of reasonable wavelengths 
that are sufficiently strong to initiate the secondary instability to the formation of persistent structures. 
Another numerical approach to determine exact coherent states tracks trajectories in the 
boundary between laminar and turbulent flows that evolve to invariant states in the boundary,
a technique known as edge tracking \cite{Skufca:2006iu}.

\begin{figure}[ht]
\includegraphics[height=0.17\textheight]{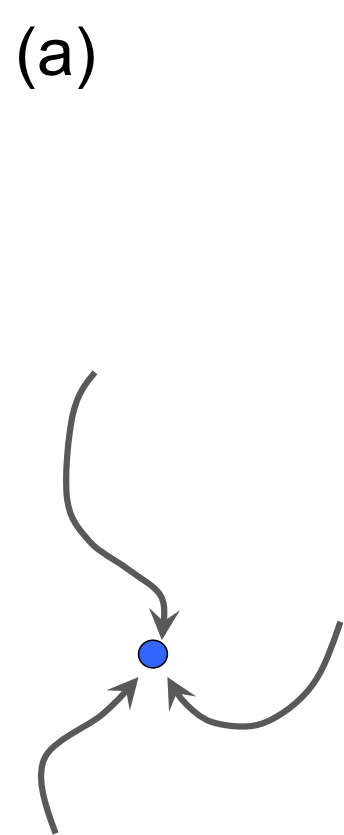}
\includegraphics[height=0.17\textheight]{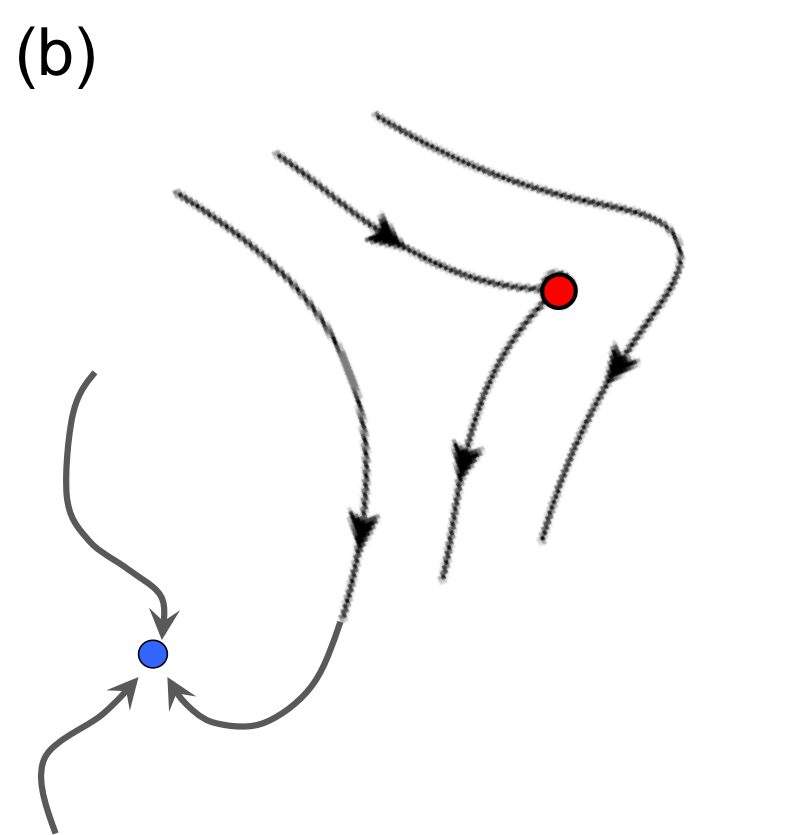}
\includegraphics[height=0.17\textheight]{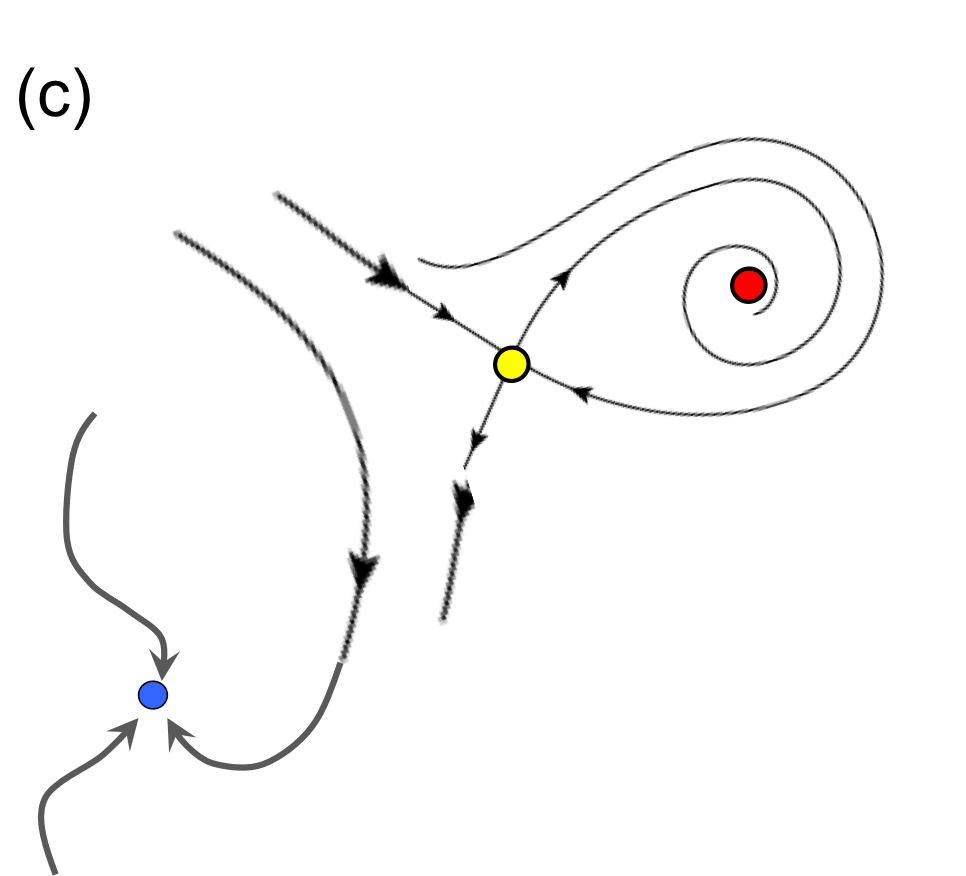}
\includegraphics[height=0.17\textheight]{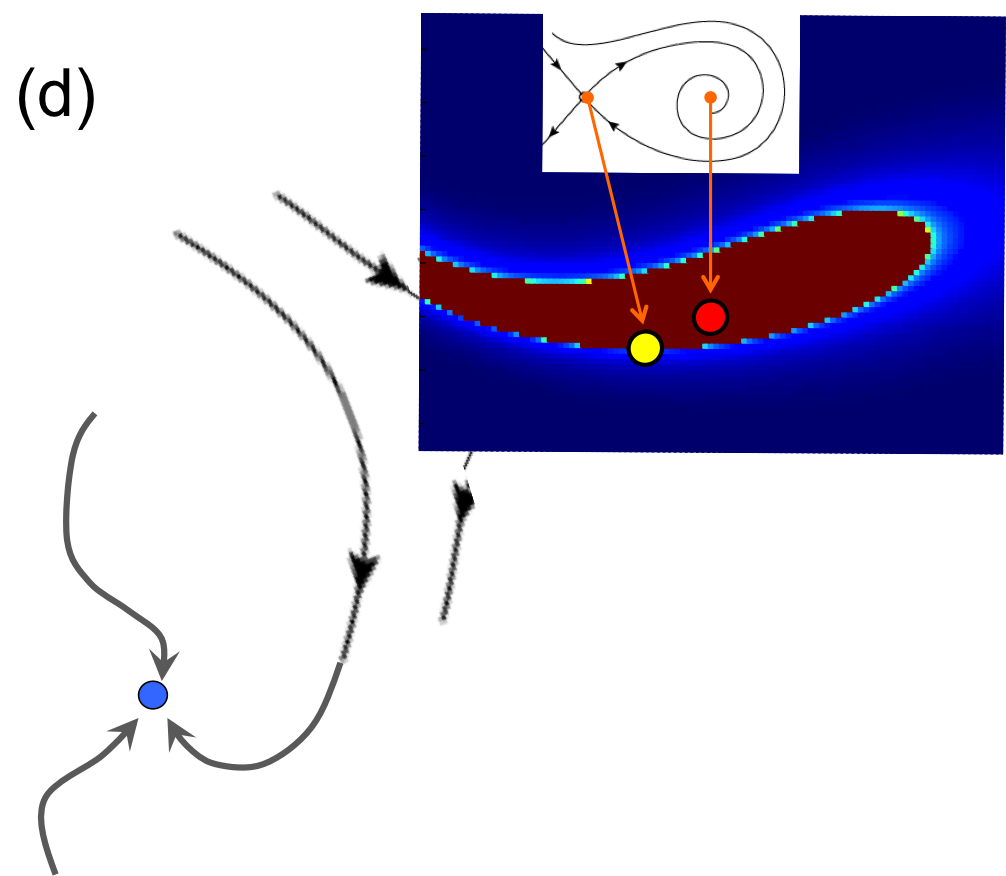}
  \caption{(Colour online) A sketch of the saddle-node bifurcation in the state-space of 
  the system. (a) Below the appearance of exact coherent structures, all initial conditions
  return to the laminar profile (blue dot), as indicated by the arrows. (b) At the critical point,
  one set of initial conditions converges to the new exact coherent structure (red dot). (c) Above the critical
  point a saddle (yellow) and a node (red) appear and a finite regions of initial
  conditions is attracted to the node. (d) The region attracted to the node can be
  mapped out by following the time evolution of initial conditions: in the blue region
  they return to the laminar profile, in the red region that approach the node. The figure
  is adapted from the results in \cite{Kreilos:2012bd}.}
  \label{fig:bif_structure}
\end{figure}

An example of the bifurcations that take place in the state space of the system 
is shown in Fig. \ref{fig:bif_structure}, for the case of plane Couette flow \cite{Kreilos:2012bd}. 
Initially, below the formation of the 
structures, all initial conditions return to the laminar profile indicated by the blue circle.
As one approaches the critical point for the onset of the structure, one initial condition
is slowed down and ends up in the critical point, indicated in red. For higher
Reynolds numbers, a saddle point (yellow) and a node (red) appear. Further away
from the saddle point the region that is attracted to the node becomes very
thin \cite{Lebovitz:2009fb}.
The saddle point has stable and unstable directions, with the unstable directions 
either pointing towards the laminar profile or to the new node. 
In the 2-d example shown the saddle has one
unstable direction, and the node has only stable directions. In higher-dimensions
that may be more complicated, and the saddle-node bifurcation may happen in
an unstable subspace. However, the saddle always has at least one
more unstable direction than the node. 

The region in state space that will be affected by this bifurcation has the shape
of a droplet, as indicated in (d): the picture is obtained by mapping out in a
2-d cross section of state space the fate of initial conditions in that plane (Ref. \cite{Kreilos:2012bd}): 
in the blue regions all initial conditions return to the laminar profile. In the red/brown area they are 
attracted to the nodal state. The tail of the droplet shaped region becomes very thin and 
reaches out to large amplitudes. As the Reynolds number increases,
the bulk region becomes wider. The secondary bifurcations take place within the
region and have little effect on its shape, until the crisis bifurcation that rips the closed
attracting region open and turns the attractor into an open chaotic saddle.

\begin{figure}[ht]
\begin{center}
\includegraphics[width=0.9\textwidth]{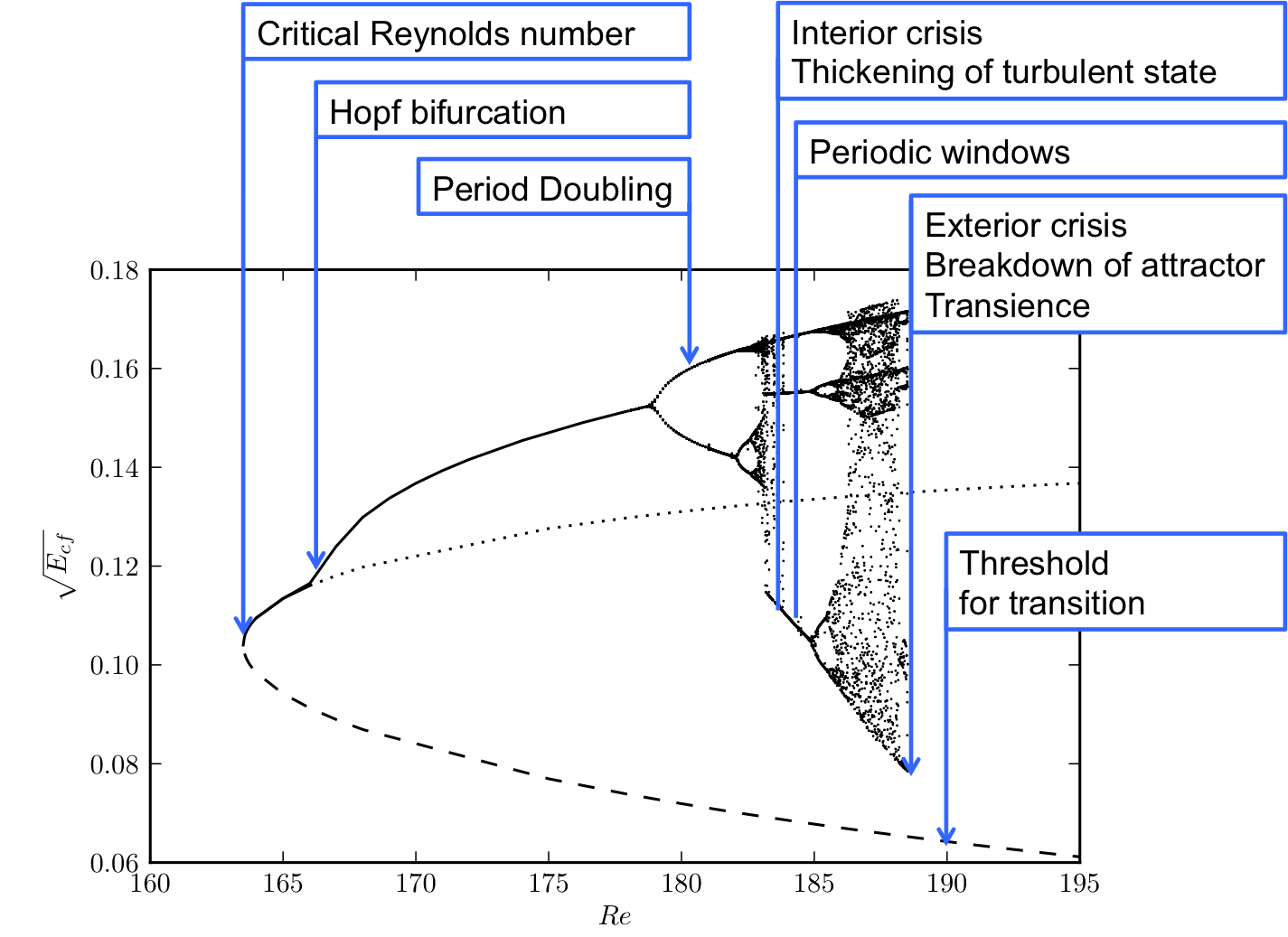}
\end{center}
  \caption{(Colour online) 
Secondary bifurcations of exact coherent structures. The exact coherent structures appear at the
critical Reynolds number, undergo a Hopf bifurcation followed by period doubling
bifurcations, various interior crises in which the attractor becomes wider
\cite{Kreilos:2014ew}, the familiar periodic windows, and eventually a crisis
bifurcation where the attractor collides with the saddle state (indicated by the dashed
line) and turns into a strange saddle. The node also determines the threshold
which perturbations have to cross in order to trigger turbulence.}
  \label{fig:bif-diagram}
\end{figure}

The nodal state undergoes secondary bifurcations that add temporal complexity
\cite{Kreilos:2012bd,Zammert:2015jg,Avila:2013jq}:
the bifurcation diagram shown in Fig. \ref{fig:bif-diagram} looks very much like
that of the logistic or quadratic map $x_{n+1} = 4 \lambda x_n (1-x_n)$ as the parameter
$\lambda$ is increased: a Hopf bifurcation brings in a frequency that turns the stationary
states (in plane Couette flow) and the travelling waves (that are stationary states
in a comoving frame of reference, in plane Poiseuille flow or pipe flow) into relative
periodic orbits. Upon further increase in Reynolds number, they undergo a sequence
of period doubling bifurcations, and then show a series of chaotic bands with
periodic windows in between. Further increasing the Reynolds number results
in a collision between the attractor that has formed around the node and a stable
direction of the saddle. In such a \textit{crisis bifurcation} \cite{Grebogi:1982ck} 
the attractor changes to a transient chaotic saddle. With the chaotic saddle comes the 
sensitive dependence on initial conditions seen in experiment \cite{Darbyshire:1995hs} and 
simulations \cite{Schmiegel:1997vf}, as well as the possibility to decay and to return to the
laminar flow \cite{Faisst:2004bs,Hof:2006ab}.

When additional parameters, such as the wavelength in the neutral directions, are taken
into account, the unfolding of the saddle-node bifurcation becomes more complex
\cite{Mellibovsky:2011gq,Mellibovsky:2012kf}.

The lack of stable coherent structures and the chaotic nature of the dynamics above onset 
make it very difficult to detect exact coherent structures in experiments. With judiciously
chosen observables, it is possible to pick them out in velocity fields and to see the transitions
between a few of these patterns \cite{Hof:2004ab,Schneider:2007ib,Kerswell:2007ds}.
Moreover, it is possible extract velocity fields from observations and to continue their evolution
numerically and to thereby verify that they stay close to invariant solutions for 
some time \cite{deLozar:2012}.


\section{Localization}
\begin{figure}
\includegraphics[width=0.95\textwidth]{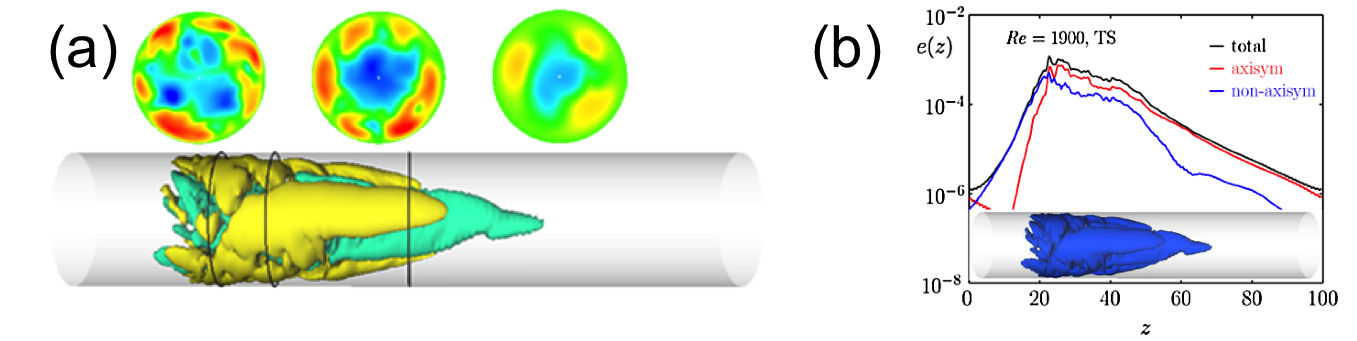}
  \caption{(Colour online) 
  A localized state in pipe flow, taken from \cite{Mellibovsky:2009fp}. (a) the highspeed
  streaks in the center (green) and the lowspeed streaks near the wall (yellow) together
  with cross sections of the pipe. (b) energy content of the puff along the
  axis of the pipe; the data are compatible with an exponential decay.
}
  \label{fig:pipe_localized}
\end{figure}

The coexistence of a turbulent state (even if it is only transient) and a stable laminar state 
provides key ingredients for the formation of localized turbulent patches with transitions
(fronts) between the laminar and turbulent state on their boundary. 
As an early example for shear flows, one finds the numerical experiment by 
Lundbladh and Johannsson \cite{Lundbladh:1991xy}:
they initiated a localized patch in plane Couette flow and studied
its evolution in time for different Reynolds numbers. If the Reynolds number was too low, the patch
would shrink and disappear. For higher $\Rey$, the patch would grow and expand in all directions.
More extensive calculations by \cite{Duguet:2010dv} showed that the patch does not stay uniform
but develops alternating oblique bands of turbulent and laminar regions. 
Studies by Barkley and Tuckerman \cite{Barkley:2005ko} obtained similar structures from homogeneous 
turbulent states upon decreasing the Reynolds number. Further examples of localized
turbulence come from observations on puffs and slugs in pipe flow \cite{Rotta:1956vt} 
(see Fig. \ref{fig:pipe_localized}) or spiral turbulence in Taylor-Couette flow \cite{Prigent:2002fq}.

The mechanisms by which a localized turbulent spot can spread differ in the spanwise
and downstream direction. In the spanwise direction studies of spots between plates
show that the longitudinal vortices at the boundary drive further rolls in their neighborhood
\cite{Duguet:2011kd,Duguet:2013vf,Schumacher:2001gb}.
This process is not monotonic, and some of the rolls that are
triggered may decay again, which is consistent with the transient nature of local structures.
In the flow direction spreading is driven by the advection of turbulent blobs that are pulled out
of the turbulent region. 

The dynamics of localized spots is not restricted to such processes: it is also possible that
the entire structure moves around. In the case of the suction boundary layer,
various forms of localized, periodically oscillating or propagating and erratically meandering
states have been found \cite{Khapko:2013hs,Khapko:2014fl,Khapko:2016ev}.
Similar structures have also been observed in plane Poiseuille flow 
\cite{Tuckerman:2014kn,Xiong:2015ei}.

Pipe flow shows an even larger variety of structures: the first structures that appear, so-called 
puffs \cite{Wygnanski:1973fp}, 
do not grow to fill the pipe but maintain a certain length (the example shown in Fig. \ref{fig:pipe_localized}).
For higher Reynolds numbers, they show a transition to another form that differs in the intensity profiles 
along the axis, so-called slugs. For slugs, the fronts move at different speeds so that they 
spread along the pipe. As more detailed theoretical and experimental studies show
\cite{Barkley:2015jn,Barkley:2016wn}, there are two kinds 
of slugs, distinguished by details of their intensity profiles and their spreading speeds.
The characterization of weak and strong slugs follows from intriguing connections to the 
nonlinear dynamics of pulse propagation on nerve cells and raises interesting questions
about frontal dynamics, as explained in Barkleys valuable perspectives article \cite{Barkley:2016wn}.

\begin{figure}
\begin{center}
\includegraphics[width=0.9\textwidth]{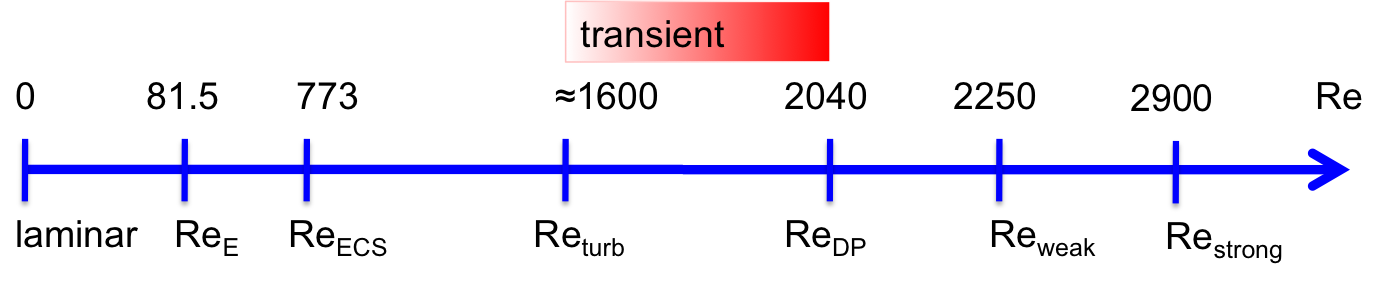}
\end{center}
  \caption{(Colour online) 
  The various Reynolds numbers of pipe flow along the Reynolds number axis
  in an updated version of the diagram from \cite{Eckhardt:2011gl} (not to scale): 
  energy stability $\Rey_E=81.5$  \cite{Joseph:1969th};
  appearance of the first critical states near $\Rey_{ECS}=773$ (Ref. \cite{Pringle:2007hz});
  first indications of turbulence in experiments near $\approx 1600$ \cite{Willis:2008bb};
  critical Reynolds number from balance between splitting and decay $\Rey_{DP}\approx 2040$ \cite{Avila:2011hq}; 
  weak spreading of slugs above $\Rey\approx 2250$; 
  and finally transition to strong spreading near $\Rey=2900$ \cite{Barkley:2015jn}.}
  \label{fig:Repipe}
\end{figure}

The relation between large scale coherent structures and the extended coherent structures 
described in the previous section is only partly understood. Studies on structures in large domains 
\cite{Duguet:2009bd,Schneider:2010id,Melnikov:2014fl,Chantry:2014ex,Mellibovsky:2015ep}
show that extended structures are susceptible to long wavelength modulations which trigger 
transitions to localized structures. The connection to snaking bifurcations 
\cite{Knobloch:2008kv,Schneider:2010jz,Gibson:2016jk}
suggests that there are entire families of localized structures with different numbers of vortices
and different lengths and widths. In the downstream direction, the decay of the structures tends
to be exponential, as shown in Refs. \cite{Gibson:2014jr,Brand:2014he,Zammert:2016dn,Ritter:2018ic} 
using approximate representations of the velocity fields. The full analysis using representations 
for incompressible flow fields
and allowing for modulations in the decay, is described in
\cite{Barnett:2017im}.
In the spanwise direction, the decay seems to be even more rapid. In two-dimensional
situations, e.g. with localized structures in plane Couette flow \cite{Schumacher:2001gb}, 
one finds large scale flow structures that decay more slowly. The decay of states is not only of
interest in its own right but also controls the interactions between patterns and therefore has
implications for the relation to non-equilibrium phase transitions that described in the next section.


\section{Spatiotemporal dynamics and directed percolation}
The coexistence of the stable laminar state and a non-persistent turbulent state allows
for spatio-temporal dynamics that is reminiscent of directed percolation, a paradigmatic
non-equilibrium phase transition \cite{Hinrichsen:2000cy}. Observations of puffs show that while
they decay after some time \cite{Hof:2006ab,Hof:2008cb}, they can also nucleate another puff 
when a turbulent blob is extracted, pulled downstream by the mean shear, and then reattached 
to the wall. If a puff is more likely to decay than to nucleate another one, the total
number of puffs will decrease until the laminar profile is restored. If the puffs will split sufficiently
frequently before they decay, then the number of puffs will increase and a finite fraction of
space will remain turbulent. As described in \cite{Avila:2011hq}, both processes can be viewed
as independent random events, with exponential distributions in the times before an event
(decay or split) occurs. Avila et al \cite{Avila:2011hq} found that the 
mean time $T_d$ for a decay increases with Reynolds number like
\beq
T_d 
\approx \exp\exp\left(\frac{\Rey-1550}{180}\right)
\eeq
and the mean time for a splitting event decays with Reynolds number like
\beq
T_s\approx \exp\exp\left(\frac{2940-\Rey}{320}\right)
\eeq
The stronger than exponential variation of life times with Reynolds number is robust
and has also been seen in other systems. To date, the most plausible explanation 
for the super-exponential variation uses a link to large fluctuations and 
rare events \cite{Goldenfeld:2010hm}, and such arguments could 
also motivate the faster than exponential variation of the time to a splitting event.
Equating the decay and splitting times gives an estimate for a Reynolds number where 
the two processes balance: $\Rey_c \approx 2040$. For $\Rey$ below $\Rey_c$ all turbulent 
patches will eventually decay and the system will return to the laminar profile. This 
definition of a critical Reynolds number can be used to replace the variety of 
critical Reynolds numbers that can be found in the literature \cite{Eckhardt:2009bb}
and it can also be used for other flows \cite{Shi:2013cg}.

\begin{figure}[ht]
\begin{center}
\includegraphics[width=0.4\textwidth]{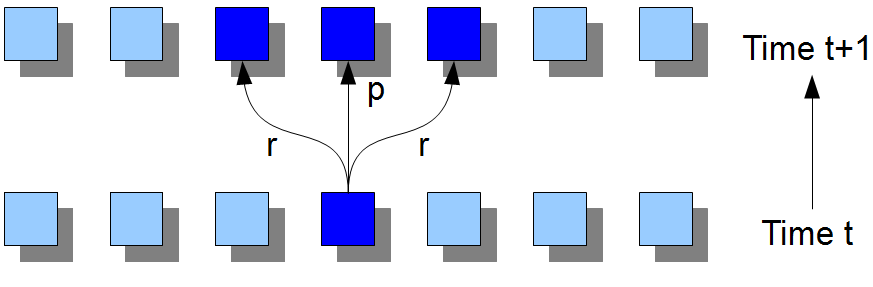}
\includegraphics[width=0.5\textwidth]{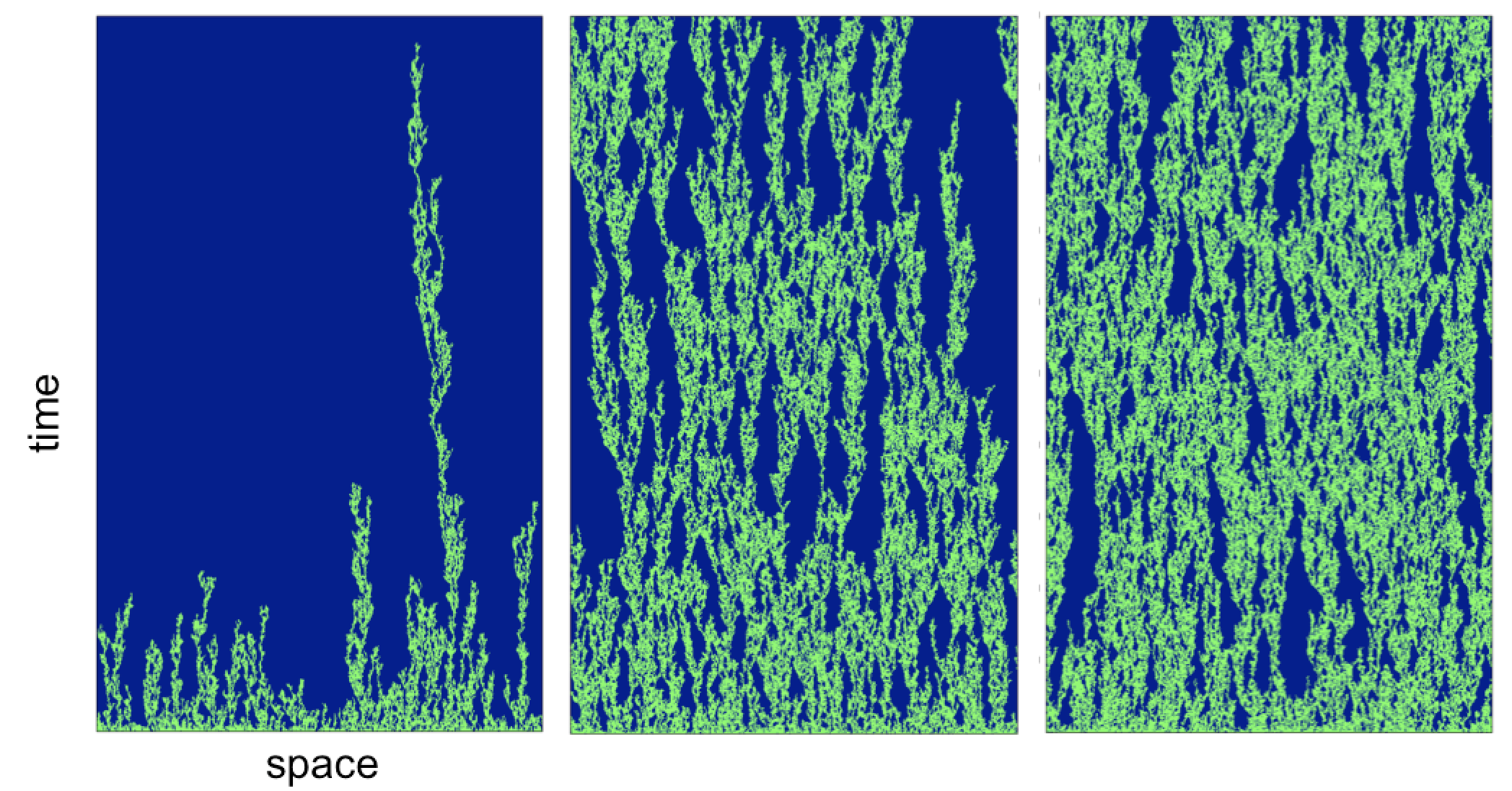}
\end{center}
  \caption{(Colour online) 
The dynamics in 1d directed percolation. (a) a turbulent cell can 
stay turbulent (with probability $p$) and introduce turbulence
in the neighboring cells (with probability $r$). (b) typical space-time diagrams
in directed percolation following the dynamics in (a). The light green areas are turbulent cells.
The first frame is below the critical point where all turbulent cells die out. The middle frame 
is at the critical point, and the right frame is above the critical point. }
\label{fig:DP_model}
\end{figure}

The dynamics of decay and splits in the turbulent puffs is reminiscent of
the dynamics in directed percolation, thereby providing a first realization of
a suggestion of Pomeau \cite{Pomeau:1986vj,Pomeau:2016cd,Manneville:2009if}.
Directed percolation is a time-discrete process on a discrete lattice: states are
labelled laminar or turbulent, and a turbulent state can remain turbulent with a certain
probability $p$ and induce transitions in neighboring cells to become turbulent 
with another probability $r$ (see Fig. \ref{fig:DP_model} for the model studied in \cite{Allhoff:2012ej}. 
Identifying a puff with the turbulent state, the decay and spreading mimick the dynamics 
in directed percolation. The Reynolds number is the control parameter and the 
critical Reynolds number marks the transition point. The turbulent fraction $f$ can
be taken as the order parameter for the transition, and it varies for 1-d percolation like
\beq
f \propto (\Rey-\Rey_c)^{0.276} \,.
\eeq
Other observables like the length and width of laminar regions scale with different exponents
\cite{Hinrichsen:2000cy}. The exponents have not been measured for pipe flow because of the excessively
large observation times. However,  the critical exponents could be measured in a thin
Taylor-Couette cell \cite{Lemoult:2016ks}, and all exponents have been found to be in good agreement
with theoretical predictions. 

In pipe flow, and also in duct flow or narrow Taylor-Couette flow, 
the turbulence is by and large homogeneous across a cross section,
so that only one degree of freedom, along the axis, has to be taken into account.
In the two-dimensional cases like plane Couette flow or plane Poiseuille flow \cite{Sano:2016kh}, 
the turbulent regions are not homogeneous, and there are observations that indicate that there
several non-laminar states \cite{tao1,tao2}. Therefore, the relation to directed percolation 
in 2d systems is less clear \cite{Pomeau:2016cd,chantry}.


\section{Conclusions and outlook}
The past decades have seen much progress in our understanding of the transition to turbulence
in shear flows without linear instabilities. The presence of shear drives vortices and streaks so
that the energy content in the perturbations is transiently amplified. The amplification shows how
shear flows become more susceptible to perturbations as the flow rate increases. Eventually,
nonlinear effects become important and new, three-dimensional states appear in subcritical
bifurcations. They provide a scaffold that can support turbulent dynamics: phrased the other way
around, no turbulence is possible until such states are present. The coexistence of the new 
states with the stable laminar profile allows for the development of spatially localized patterns,
such as the puffs and slugs in pipe flow.

Initially, the regions in state space that are affected by the new states are small, 
but as ever more states appear, they connect and entangle to form a web that can sustain 
turbulence for long times \cite{Halcrow:2009jx,Gibson:2016jk,avilaNJP}. 
In spatially extended systems a variety of patterns can arise,
with complex dynamics between different patterns and on the boundary to the laminar profile.
The coexistence of the turbulent dynamics with a stable laminar profile suggests links
to the non-equilibrium phase transition of directed percolation, and for 1d systems there is
evidence that the transition has all the characteristics of directed percolation. It remains
a challenge to establish similar results for transition in 2d flows. 

One way to summarize the different processes and transitions is the sequence of Reynolds numbers
shown in Fig. \ref{fig:Repipe}. It will be interesting to explore similar sequences and processes
in other flows.

All the examples described here are parallel shear flows that are translationally invariant in the
downstream direction. For spatially developing flows, such as boundary layers, the local
variation in Reynolds numbers has to be taken into account. Some of the concepts described
above can be transferred also to boundary layers \cite{Cherubini:2011dk,Duguet:2012kb}, 
and steps towards the explanation of the
spatiotemporal fluctuations near the transition \cite{Emmons:1958tj,Narasimha:1985ui}
have been taken \cite{Cherubini:2010cp,Vinod:2004ca,Vinod:2007vd,Kreilos:2016gq}.

For higher Reynolds numbers beyond the transition to turbulence, the challenging question is
the emergence of the turbulent cascade. For this, the dynamics has to become not only
temporally but also spatially complex . It will be interesting to see to which extend the ideas developed
for the transition can be carried over to the statistical mechanics of the turbulent cascade.

\section*{Acknowledgement}
This work has been supported in part by the Deutsche Forschungsgemeinschaft within its
priority program "Turbulent superstructures".

\begin{small}

\end{small}

\end{document}